\documentclass[preprint,12pt,number]{elsarticle}

\usepackage{multirow,makecell}
\usepackage{tcolorbox}
\usepackage{color,xcolor}
\usepackage{listings,amsfonts}
\usepackage{caption}
\usepackage{subcaption}
\usepackage{threeparttable}
\usepackage{bbding}
\usepackage{graphicx}
\usepackage{booktabs}
\usepackage{tabularx}
\usepackage{longtable}
\usepackage{enumitem}
\usepackage{array}
\usepackage{framed}
\usepackage{color}      
\definecolor{shadecolor}{rgb}{0.95,0.95,0.96}

\usepackage{hyperref}

\definecolor{customcite}{HTML}{b83b5e}
\definecolor{customlink}{HTML}{07689f}
\definecolor{customurl}{HTML}{11999e}

\hypersetup{
  linkcolor = customlink,
  citecolor = customcite,
  urlcolor = customurl
}

\newcommand{\tool}{\textsc{LaQ\-ual}}

\usepackage{url}

\usepackage{amssymb}

\usepackage{amsmath}

\journal{Journal of Systems and Software}

\begin{document}

\begin{frontmatter}

\title{\tool{}: An Automated Framework for LLM App Quality Evaluation}

\author[inst1]{Yan Wang}
\author[inst2]{Xinyi Hou}
\author[inst1]{Junjun Si\corref{cor1}}
\author[inst2]{Yanjie Zhao} 

\author[inst1]{Weiguo Lin}
\author[inst2]{Haoyu Wang}

\cortext[cor1]{The corresponding author is Junjun Si (sijunjun@cuc.edu.cn).}
\affiliation[inst1]{organization={School of Computer and Cyber Security, Communication University of China},
            city={Beijing},
            country={China}}
\affiliation[inst2]{organization={School of Cyber Science and Engineering, Huazhong University of Science and Technology}, 
            city={Wuhan}, 
            country={China}}

\begin{abstract}
Representing a new paradigm in software distribution, LLM app stores are rapidly emerging, offering users diverse choices for content generation, coding assistance, education, and more. However, current ranking and recommendation mechanisms in LLM app stores predominantly rely on static metrics, such as user interactions and favorites, making it challenging for users to efficiently identify high-quality apps. At the same time, current academic research focuses on specific vertical fields and lacks a general, automated evaluation framework applicable to the diverse LLM app ecosystem. To address the above challenges, we present \tool{}, an automated framework for \underline{\textbf{L}}LM \underline{\textbf{a}}pp \underline{\textbf{qual}}ity evaluation. \tool{} integrates three key stages: (1) LLM app labeling and hierarchical classification for precise scenario mapping; (2) static indicator evaluation using time-weighted user engagement and functional capability indicators to filter low-quality apps; and (3) dynamic scenario-adaptive evaluation, where an LLM generates scenario-specific evaluation metrics, scoring criteria, and tasks for comprehensive quality evaluation. Experiments on a mainstream LLM app store demonstrate the effectiveness of \tool{}. Its automated scores show high consistency with human judgments (Spearman's $\rho$ = 0.62, $p=0.006$ in legal consulting; $\rho$ = 0.60, $p=0.009$ in travel planning). Through effective screening, \tool{} can reduce the candidate LLM app pool by 66.7\% to 81.3\%. User studies further validate its significant outperformance over baseline systems, particularly in comparison efficiency (mean 5.45 vs. 3.30) and value of explanatory information (4.75 vs. 2.25). These results demonstrate that \tool{} provides a scalable, objective, and user-centric solution for high-quality discovery and recommendation of LLM apps in real-world scenarios.

\end{abstract}


\end{frontmatter}

\section{Introduction}
\label{Introduction}

Large language models (LLMs) have made breakthrough progress in recent years, which has spurred the emergence of a wide range of LLM-based intelligent software, such as LLM apps, LLM agents, self-hosted LLM services, and edge-LLM applications. Among these software forms, this study focuses on \textbf{LLM apps}~\cite{zhao2025llm}, which specifically refer to intelligent applications built upon LLM capabilities and available on various \textbf{LLM app stores} such as OpenAI's GPT Store~\cite{openaigpts}, Quora's Poe~\cite{poe}, Baidu's Smart Cloud Qianfan AppBuilder (AppBuilder)~\cite{baiduapp}, and Tencent's Yuanqi Platform~\cite{yuanqiagent}. These LLM apps typically feature independent entry points, functional descriptions, and user interfaces, and are widely used in scenarios such as text generation, content creation, education, coding assistance, and emotional support~\cite{zhao2025llm,poe,coze,cici}. Compared to traditional apps, LLM apps not only understand and generate natural language but also support complex capabilities such as external knowledge base integration, plugin invocation, and multimodal input, significantly expanding the boundaries of artificial intelligence (AI) applications~\cite{Yan2024Exploring,yan2025understanding,wu2024an,Hou2025On}.
Driven by rapid technological advances and lowered development barriers, the LLM app ecosystem has experienced explosive growth, with mainstream LLM app stores hosting a vast number of LLM apps. In just a few months since its launch, the GPT Store has already included over 3 million GPT apps~\cite{openai2023gpts}, FlowGPT has over 4 million monthly active users~\cite{pr2024flowgpt}, and platforms like Poe, and HuggingChat have each exceeded 10 million monthly visits~\cite{similarweb_poe,similarweb_huggingface}. Furthermore, third-party platforms such as GPTStore.AI~\cite{gptstoreai} and GPTs Hunter~\cite{gptshunter} have also emerged, with the app inventory on a single platform reaching hundreds of thousands, demonstrating explosive growth in overall market size and activity.

\begin{figure}[t!]
    \centering
    \includegraphics[width=\linewidth]{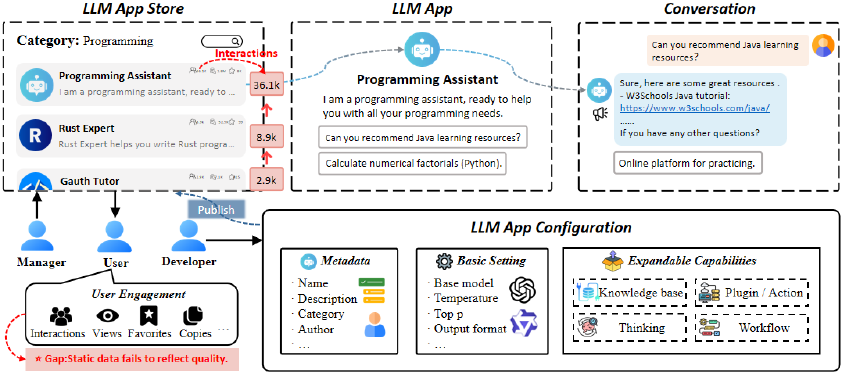}
    \caption{Overview of the LLM App Ecosystem.}
    \label{fig: Core Elements of an LLM Application}
\end{figure}

This rapid expansion brings new challenges for users in discovering and selecting high-quality LLM apps. On the one hand, the surge in the number of apps has led to information overload, making it difficult for users to find high-quality apps that truly meet their needs efficiently. On the other hand, current mainstream LLM app stores generally adhere to traditional recommendation logic: users input searching keywords, and the platform then returns a list of apps ranked by static metrics such as usage, number of favorites, and release date~\cite{Chen2016Mobile,Maqbool2023MobileRec}. With this scheme, users must manually filter and repeatedly try out various apps, incurring significant cognitive and time costs. In addition, recent user discussions have revealed that app rankings are frequently manipulated through methods such as keyword stuffing, fake reviews, and misleading updates, resulting in low-value apps appearing at the top of search results~\cite{fourm1,fourm2,fourm3}. In summary, as illustrated in Figure~\ref{fig: Core Elements of an LLM Application}, the current mechanism, which relies on static metrics, makes it difficult for LLM App users to accurately identify truly high-quality apps, thus severely damaging the user experience and the health of the ecosystem.

Although LLM evaluation methods evolve, they typically focus on specific domains, such as law, mathematics, or code generation, and rely on corresponding benchmarks with predefined metrics~\cite{zheng2023codegeex,wang2024mmlu,NEURIPS2024_6c1d2496,guha2023legalbench,rao2025software}. While these benchmarks are effective for evaluating LLMs in well-defined scenarios, it's difficult to adapt them to the LLM app store context. The diversity of app categories can reach dozens or even more, making a comprehensive evaluation of various types of apps impractical using existing methods~\cite{su2025gpt,zhang2024lookgptappslandscape}. Furthermore, unlike general-purpose or domain-specific LLMs, individual LLM apps often target highly specialized or niche functionalities, which require evaluation at a much finer granularity to accurately reflect their value and effectiveness. Most importantly, the ultimate goal of LLM app stores is to serve users; therefore, factors such as user engagement, experience, and satisfaction must be fully considered in the evaluation process. However, the above benchmarks are mostly model-centric and static, lacking mechanisms to capture real user interactions or feedback. Therefore, a new evaluation framework that can adapt to the multidimensional, fine-grained, and user-centric nature of the LLM app ecosystem is urgently needed.

In response to these unmet needs, we propose \tool{}, an automated framework for \underline{\textbf{L}}LM \underline{\textbf{a}}pp \underline{\textbf{qual}}ity evaluation, designed to be broadly applicable across diverse LLM app scenarios. \tool{} first conducts scenario-aware labeling and classification to accurately map LLM apps to their intended use cases. Next, it performs multi-dimensional static filtering based on time-weighted user engagement and functional capability indicators to efficiently exclude low-quality candidates. Finally, \tool{} employs LLM-driven dynamic assessment, automatically generating scenario-adaptive evaluation metrics and tasks to enable fine-grained automatic LLM app quality measurement.

We systematically evaluate the effectiveness of our framework from three key perspectives: evaluation accuracy and adaptivity, recommendation  effectiveness, and user impact. First, we assess whether \tool{} can provide accurate and scenario-adaptive quality evaluation by comparing its automated scores with human judgments of 28 human evaluators (including domain experts) across different LLM app scenarios. Second, we benchmark \tool{} against mainstream app store recommendation mechanisms, demonstrating that our framework substantially reduces the candidate app pool while surfacing higher-quality apps. Third, we examine the impact of \tool{} on users' decision quality, cognitive effort, and trust in real-world app discovery tasks through a controlled user study with 12 participants ($N=12$) using a within-subjects design.
Results show that \tool{} improves decision support and comparison efficiency for users, and increases the perceived value of explanatory information compared to baseline systems.
Collectively, these evaluations validate the practical effectiveness, rationality, and user-centered benefits of our solution in real-world LLM app scenarios.

In summary, this work makes the following key contributions:

\begin{itemize}
\item We design and implement \tool{}, an automated framework featuring a multi-stage pipeline for scalable and systematic LLM app quality evaluation, achieving both efficient and accurate evaluation results.

\item We propose an efficient static indicator evaluation module that utilizes time-weighted user engagement and functional capability indicators, reducing low-quality app candidates by 66.7\%–81.3\% and significantly improving evaluation efficiency.

\item We develop a fully automated dynamic evaluation pipeline, where LLMs are employed to generate scenario-specific evaluation metrics and tasks, enabling fine-grained, context-adaptive, and objective quality evaluation across diverse app categories, achieving strong consistency with human judgments (Spearman’s $\rho = 0.62$, $p=0.006$ in legal consulting, and $\rho = 0.60$, $p=0.009$ in travel planning). 

\item We conduct comprehensive real-world validation, showing that \tool{} achieves high agreement with expert ratings, and substantially boosts user decision confidence (mean 5.05 vs. 4.40), comparison efficiency (5.45 vs. 3.30), and perceived value of explanatory information (4.75 vs. 2.25) compared to baseline systems. 

\end{itemize}

The remainder of this paper is organized as follows. \autoref{sec:background} reviews relevant background on LLM apps and LLM-based evaluation. \autoref{sec:framework} introduces the architecture and key components of \tool{}. In \autoref{sec:evaluation} validates the framework’s effectiveness through experiments and user studies. \autoref{sec:implication} discusses the main findings and their broader significance. \autoref{sec:limitation} outlines current limitations and future directions. \autoref{sec:related} summarizes related work, and \autoref{sec:conclusion} concludes the paper.

\section{Background}
\label{sec:background}
\subsection{LLM App}

\textbf{LLM apps} are specialized applications built on LLMs that provide intelligent, task-oriented services through natural language interaction~\cite{zhao2025llm}. Unlike traditional software, LLM apps leverage the reasoning, comprehension, and generation abilities of foundation models to support diverse scenarios, such as content creation, code assistance, consultation, translation, and data analysis. LLM apps include metadata like app name, description, instructions, categories, developer details, and sample conversations, helping users understand each app’s core features and use cases.

\textbf{LLM app stores} are digital platforms designed for the aggregation, distribution, and management of various LLM apps. Notable examples include GPT Store~\cite{openaigpts}, Coze~\cite{coze}, and Poe~\cite{poe}. These platforms provide developers with efficient tools for publishing, configuring, and hosting LLM apps, as well as supporting streamlined creation, deployment, and maintenance workflows~\cite{zhao2025llm}. Recalling the structural components in Figure~\ref{fig: Core Elements of an LLM Application}, developers can incorporate functionalities such as knowledge bases, plugins, and thinking, while enriching app metadata to improve clarity and usability.
For users, LLM app stores offer intuitive interfaces to search, browse, and discover apps by categories or tags. Users can interact with selected apps through multi-turn natural language conversations, experiencing a variety of intelligent services. The platforms also collect and utilize user behavioral data and feedback, including ratings, usage statistics, and number of conversations, as important factors in app recommendation and ranking.

In recent years, the low technical barrier for LLM app development has led to an explosive growth in the number of available apps on these platforms. This rapid proliferation creates significant challenges for users, who often struggle to efficiently identify and select high-quality apps that best meet their needs. Addressing this challenge and enabling effective discovery and evaluation of LLM apps is a critical motivation for our work.

\subsection{LLM-based Evaluation Paradigms}

Traditional evaluation methods often lack flexibility and semantic understanding when assessing complex language outputs. To address this, researchers have begun using LLMs as evaluators, a paradigm known as \textbf{LLM-as-a-Judge}. This approach leverages the LLM’s reasoning abilities to overcome the rigidity, limited scalability, and semantic shortcomings of traditional methods.
A line of pioneering studies has demonstrated the potential of this paradigm. For example, Chiang et al.~\cite{chiang2023can} showed that LLM-based evaluations are often highly consistent with expert human judgment, especially in open-ended tasks like story generation. In the education domain, Chang et al.~\cite{yang2024evaluating} and Grévisse et al.~\cite{grevisse2024llm} leveraged GPT-family and Gemini models, respectively, for automatic grading of short-answer questions, achieving promising results in terms of accuracy and efficiency compared to human annotators. These advances indicate that LLM-as-a-Judge could substantially reduce labor costs and improve handling of edge cases in various vertical applications.

Nevertheless, a growing body of research also reveals significant challenges in generalizing this approach. As summarized by Gu et al.~\cite{gu2024survey} and others, the reliability and stability of LLM-based evaluation are often sensitive to prompt design and model sampling parameters. Furthermore, many current studies rely on limited or single-source test sets and lack systematic scoring standards, which makes it difficult to extend their findings to diverse, real-world LLM application scenarios involving complex interactions.

Motivated by these gaps, this work presents the \tool{} framework, which systematically optimizes the LLM-as-a-Judge paradigm and, for the first time, applies it to the comprehensive evaluation of complex LLM apps. To overcome the core limitations of previous approaches, particularly the lack of explicit evaluation criteria and insufficiently diverse test sets, \tool{} introduces a dynamic generation mechanism that automatically produces task-specific scoring metrics and evaluation questions before assessment.

\section{Proposed framework}
\label{sec:framework}

\begin{figure}[htbp]
    \centering
    \includegraphics[width=\linewidth]{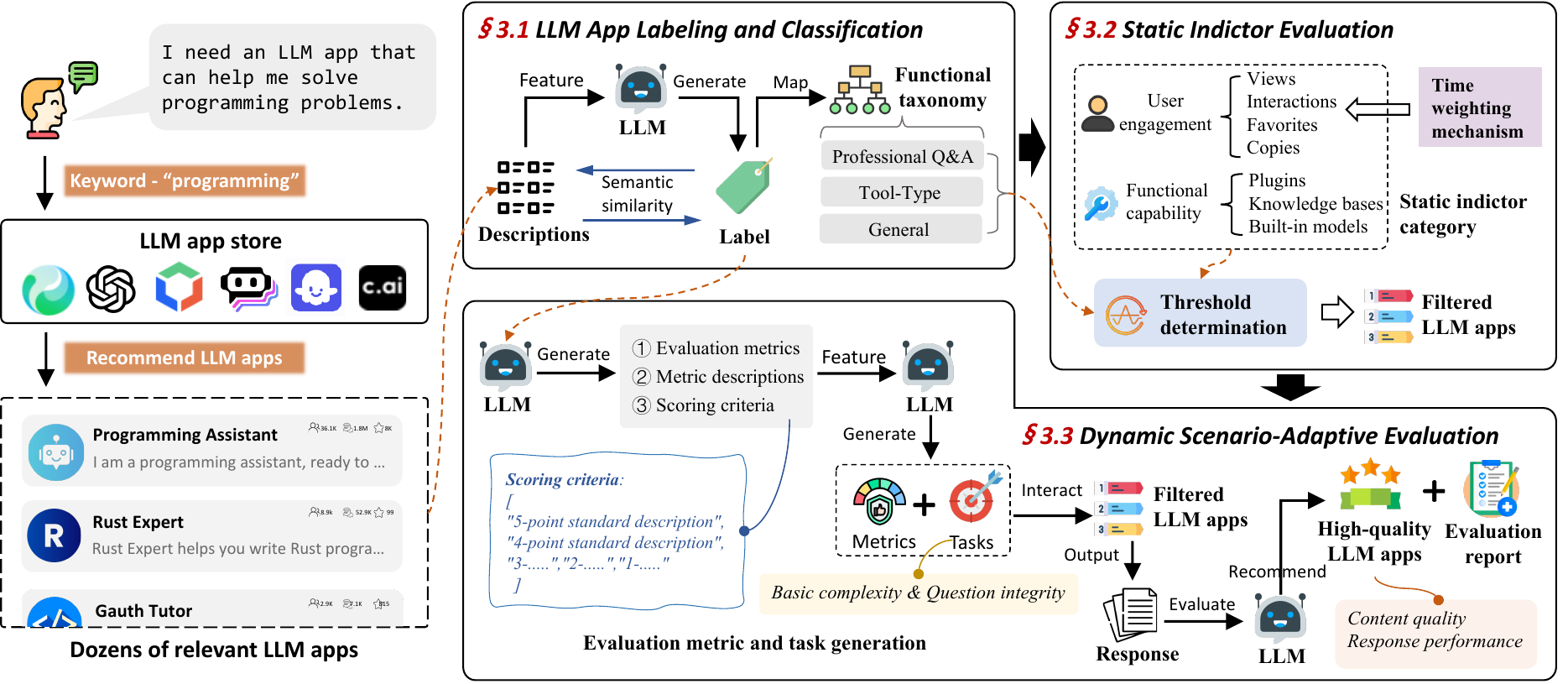}
    \caption{The Workflow of \tool{}.}
    \label{fig: Framework Diagram}
\end{figure}

This section introduces the workflow of \tool{}. As shown in Figure~\ref{fig: Framework Diagram}, \tool{} consists of three key stages: \textit{LLM app labeling and classification} (\autoref{subsec:tag}), \textit{static indicator evaluation} (\autoref{subsec:static}), and \textit{dynamic scenario-adaptive evaluation} (\autoref{subsec:dynamic}).

\subsection{LLM App Labeling and Classification}
\label{subsec:tag}
This subsection is the foundation of \tool{}, transforming unstructured metadata into standardized semantic representations for categorization. Based on raw app descriptions, the system automatically generates labels via an LLM-based annotator. These labels are then mapped to a hierarchical taxonomy, ensuring a basis for the subsequent evaluation.

\subsubsection{LLM-based Labeling}

\begin{figure}[htbp]
    \centering
    \includegraphics[width=\linewidth]{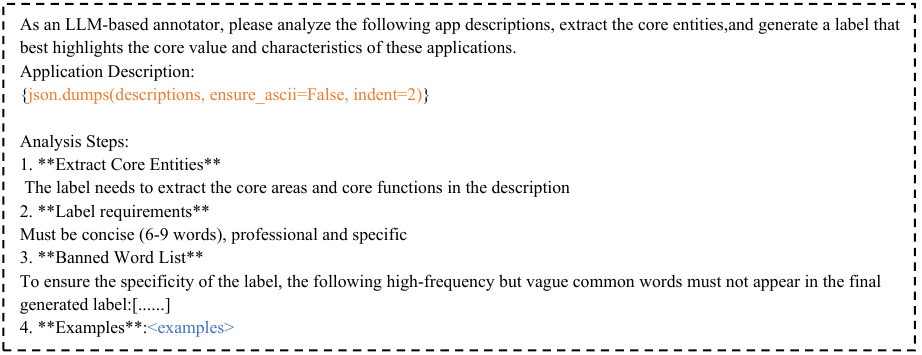}
    \caption{Prompt Template for LLM-based Annotator.}
    \label{fig:prompt_1}
\end{figure}

Before the formal evaluation begins, an \textit{LLM-based annotator} automatically generates labels for all LLM apps to be evaluated. In LLM app stores, each app is associated with a description. The LLM-based annotator performs semantic analysis on this textual information to extract key domain entities, functional elements, and typical usage scenarios. Based on this analysis, the annotator generates a candidate label that summarizes the core functionality of the app. To ensure that labels are concise and accurate, \tool{} limits their length. For example, if it is Chinese, it can be limited to a phrase containing 6-9 Chinese characters.
Figure~\ref{fig:prompt_1} illustrates the prompt of the LLM-based annotator.

We then use the \textit{text2vec} model to further verify the accuracy of the generated labels. Specifically, this model calculates the semantic similarity between each candidate label and the original description of the LLM app. A similarity threshold of 0.7 is set; if the calculated similarity falls below this threshold, a feedback mechanism is immediately triggered, prompting the LLM to iteratively refine the label until the similarity requirement is met or exceeded. It is important to note that the threshold of 0.7 is empirically determined, based on extensive experimental validation and a comprehensive assessment of both label quality and iteration efficiency under different thresholds. 

To avoid ambiguity, we explicitly distinguish between \emph{labels} and \emph{tags} in \tool{}. A \emph{label} is a concise semantic summary automatically generated from the app description, whereas a \emph{tag} is a predefined taxonomy item used in the subsequent hierarchical classification stage.

\subsubsection{Hierarchical Classification}
\label{subsubsec:classification}
After generating labels for each LLM app, the next step is to determine its specific category. 
As illustrated in Table~\ref{tab:llm_app_classification}, we designed a three-level hierarchical taxonomy, consisting of \textit{category}, \textit{subcategory}, and \textit{tag}, to categorize LLM apps based on their core functional characteristics.

\begin{table}[htbp]
\centering
\caption{Functional Taxonomy of LLM Apps.}
\label{tab:llm_app_classification}
\resizebox{1\linewidth}{!}{
\begin{tabular}{l l >{\raggedright\arraybackslash}p{11cm}} 
\toprule[1.2pt]
\textbf{Category} & \textbf{Subcategory} & \textbf{Tags} \\ 
\midrule[1.2pt]

\multirow{9}{*}{Professional Q\&A}
    & Medical Health & Medical, Health, Disease, Diagnosis, Treatment, Psychology\\ \cmidrule{2-3}
    & Legal Consultation & Law, Regulation, Contract, Litigation \\ \cmidrule{2-3}
    & Financial Management & Finance, Financial Management, Investing, Stocks \\ \cmidrule{2-3}
    & \multirow{2}{*}{Education} & Education, Training, Courses, Mathematics, Chinese, \\
    & & Biology, Physics, Chemistry, History, Geography, Politics \\ \cmidrule{2-3}
    & \multirow{3}{*}{Language Learning} & Translation, Vocabulary, Translation, Grammar, Language, \\
    & & Chinese, English, Japanese, German, Korean, French, \\
    & & Russian, Italian \\ \cmidrule{2-3}
    & Professional Consulting &  Consulting, Consultant, Expert \\ 
\midrule

\multirow{5}{*}{Tool-type}
    & \multirow{2}{*}{Development Tools} & Code, Programming, Development, Debugging, Algorithms,\\
    & & Front-end, Back-end, Python, Java, Javascript, Matlab \\  \cmidrule{2-3}
    & Analysis Tools & Data Analysis, Statistical Analysis, Data Forecasting \\ \cmidrule{2-3}
    & Planning Tools & Planning, Planner, Planning Design\\ \cmidrule{2-3}
    & Design Tools & Interior Design, Branding\\ 
\midrule

General
    & -- & -- \\

\bottomrule[1.2pt]
\end{tabular}
}
\end{table}

The taxonomy was constructed using a mixed approach that combined abstraction informed by prior work with corpus-based induction. Specifically, we used the classification results reported in Su et al.~\cite{su2024gpt} as an initial reference and manually abstracted candidate functional categories. Semantically similar or overlapping categories were then iteratively compared and merged through multiple rounds of refinement, resulting in 10 subcategories and 3 top-level categories: \textit{professional} Q\&A, \textit{tool-type}, and \textit{general}. For the fine-grained layer, the 62 tags were derived from high-frequency functional, usage-scenario, and domain-specific terms extracted from the names, descriptions, and related metadata of more than 2,000 apps collected from the Coze and Baidu Qianfan platforms. These candidate tags were then further refined through duplicate removal, synonym merging, and ambiguity filtering.

Rather than directly mapping the app descriptions to the predefined classification tags, \tool{} instead maps these tags using the generated labels in the previous step. \textbf{These labels serve as a semantic bridge, enabling more accurate and robust mapping to our fine-grained taxonomy by reducing the noise and ambiguity of the original descriptions.}
\tool{} employs a hierarchical matching strategy, from fine-grained to coarse-grained, prioritizing tag over subcategory. In particular, \tool{} first attempts to match label with the most specific tag keywords, such as ``litigation'' and ``contract'' under the ``Legal Consultation'' subcategory, to ensure precise classification. If no match is achieved at the tag level, the process falls back to broader subcategory keywords, such as ``Education'' or ``financial management'', for subsequent matching. 
This hierarchical retrieval approach not only ensures classification accuracy but also significantly improves the efficiency and robustness of the identification process.

\subsection{Static Indicator Evaluation}
\label{subsec:static}
As the preliminary screening layer of \tool{}, this subsection evaluates LLM apps through multi-dimensional static platform data. This process first identifies key interaction and capability indicators, then applies a time-decay weighting mechanism to ensure fairness across apps with different release dates.

\subsubsection{Static Indicator Selection}
The first step in static indicator evaluation is to determine the scope of indicators to consider. We selected two-dimensional evaluation indicators based on objective and readily available platform data. Our goal is to provide a comprehensive evaluation that reflects both user interaction patterns and the inherent technical capabilities of each LLM app.

\textbf{User engagement indicators}, which measure the app's popularity and recognition among the user community; and \textbf{functional capability indicators}, which assess the app's configuration richness and versatility. This dual-perspective approach enables us to distinguish between apps that attract broad user attention and those that offer advanced technical features. Table~\ref{tab:static_metrics} lists the detailed indicators used in our evaluation framework. For example, \textit{copies}, or configuration replication frequency, reflects an app's influence among developers, while \textit{plugins} indicates its extensibility and the ability to integrate with external tools or services.

\begin{table}[htbp]
    \centering
    \caption{Static Evaluation Indicator for LLM Apps.}
    \label{tab:static_metrics}
    \resizebox{1\linewidth}{!}{
    \begin{tabular}{ccl}
        \toprule[1.2pt]
        \textbf{Category} & \textbf{Indicator} & \textbf{Definition} \\
        \midrule[1.2pt]
        \multirow{4}{*}{User Engagement}
            & Views      & Number of times the app's main page is viewed by users. \\
            & Interactions      & Number of conversions or interactions with the app. \\
            & Favorites  & Number of times users add the app to their personal favorites. \\
            & Copies     & Number of times users duplicate the app’s configuration. \\
        \midrule
        \multirow{3}{*}{Functional Capability}
            & Plugins   & Number of external services integrated into the app. \\
            & Knowledge Bases  & Number of distinct knowledge bases connected to the app. \\
            & Built-in Models   & Number of built-in models in the app. \\
        \bottomrule[1.2pt]
    \end{tabular}
    }
\end{table}

\subsubsection{Time weighting and indicator scaling}
\label{subsubsec:time}

After determining the static indicators, we observe that newer apps in the LLM app store are often disadvantaged in evaluation because they have had less time to accumulate indicator values, whereas older apps may appear artificially strong due to the accumulation of historical data. This temporal accumulation bias can therefore obscure the real-time competitiveness of apps. To address this issue, we introduce a \textbf{time-decay dynamic weighting mechanism} that adjusts static indicators according to each app's operational age. The core idea is to ensure that the contribution of any cumulative indicator decreases as time progresses, thereby placing greater emphasis on recent performance rather than historical accumulation.

We first segment the lifecycle of each app into discrete evaluation periods, using the quarter as the basic analysis unit. Let $M$ denote the total number of months since an app’s release; the corresponding number of operating quarters $Q$ is computed as:
\begin{equation}
Q = \max(1, \lfloor M/3 \rfloor) \quad
\end{equation}
where $\lfloor \cdot \rfloor$ represents the floor function, $M$ is the total operation months and $\max(1, \cdot)$ ensures that even apps with less than three months of operation are counted as having completed one quarter. This guarantees that both new and long-standing apps are fairly included in the evaluation.

Next, for each of the user engagement indicators (i.e., views, interactions, favorites, and copies), we sequentially calculate its quarterly average and apply a time-based decay coefficient. We let \(X_{\text{total}}\) denote the total cumulative value of any given user engagement metric recorded over its entire operational period.
 
The standardized, time-weighted value $X_{\text{weighted}}$ is given by:
\begin{equation}
X_{\text{weighted}} = \left( \frac{X_{\text{total}}}{Q} \right) \cdot \omega_t \quad 
\end{equation}
This formula first calculates the quarterly mean of the indicator ($X_{\text{total}} / Q$) to eliminate the total amount difference caused by the operating time. Then, it reflects the timeliness by multiplying by a decay coefficient ($\omega_t$) that decreases over time.
The decay coefficient $\omega_t$ is defined as an exponential function of the number of operating quarters $Q$:
\begin{equation}
\omega_t = \beta^{Q-1} \quad
\end{equation}
where $\beta \in (0, 1)$ is the decay base that determines the rate at which past performance is discounted. In our implementation, we set $\beta = 0.99$, meaning each additional quarter reduces the weight of historical data to $99\%$ of the previous period. For example, a new app ($Q = 1$) retains full weight ($\omega_t = 1$), while an app active for four quarters ($Q = 4$) is weighted by $\omega_t = 0.99^3 \approx 0.9703$.

\textbf{This exponential decay mechanism effectively suppresses the inertia of historical data, thereby making the evaluation results more responsive to recent user feedback and app store dynamics.} As a result, \tool{} significantly improves both the timeliness and predictive power of the evaluation process.
The decay base $\beta$ serves as a critical hyperparameter that governs the rate at which historical data are discounted. The choice of $\beta$ should be tailored to the specific characteristics of the app store, such as platform popularity, the total number of available apps, and the level of user activity. A smaller value of $\beta$ leads to a faster decay, placing greater emphasis on the most recent market performance. In contrast, a $\beta$ value closer to one retains more of the historical performance, favoring long-term stability. By fine-tuning $\beta$, \tool{} can be flexibly adapted to different evaluation scenarios and platform characteristics.

\subsubsection{Threshold Determination}
Building upon the LLM app classification in \autoref{subsubsec:classification} and the time-weighted static indicator evaluation in \autoref{subsubsec:time}, this step focuses on the scientific determination of admission thresholds for each evaluation indicator. Once each LLM app is categorized, we establish a set of \texttt{Category-Indicator-\allowbreak Threshold} mapping rules to ensure that the entry criteria are both differentiated by category and grounded in empirical data.
Using the AppBuilder as a representative platform, we set baseline thresholds for key indicators across different LLM app categories, as shown in Table~\ref{tab:differentiated_thresholds_eng}. For example, \textit{professional} Q\&A apps are required to meet a minimum knowledge base size ($\geq 1$), while \textit{tool-type} apps are primarily assessed by the number of functional components ($\geq 2$). These thresholds are designed to guarantee that candidate LLM apps possess the essential capabilities, user engagement, and professionalism expected for their respective categories.

\begin{table}[htbp]
\centering
{
\small
\caption{Differentiated Admission Thresholds for Different App Categories (AppBuilder).}
\label{tab:differentiated_thresholds_eng}
}
\resizebox{0.9\linewidth}{!}{
\begin{threeparttable}
\begin{tabular}{llccc}
\toprule[1.2pt]
\multicolumn{2}{l}{\textbf{Indicator Dimension}} & \textbf{Professional Q\&A} & \textbf{Tool-type} & \textbf{General} \\
\midrule[1.2pt]
\multirow{4}{*}{User Engagement\tnote{1}} & Views & $\geq$20 & $\geq$20 & $\geq$20 \\
 & Interactions & $\geq$20 & $\geq$20 & $\geq$20 \\
 & Favorites & $\geq$20 & $\geq$20 & $\geq$20 \\
 & Copies & $\geq$20 & $\geq$20 & $\geq$20 \\
\cmidrule{2-5}
\multirow{3}{*}{Functional Capability\tnote{2}} & Plugins & $\geq$0 & $\geq$2 & $\geq$0 \\
& Knowledge Bases & $\geq$1 & $\geq$0 & $\geq$0 \\
& Built-in Models & $\geq$1 & $\geq$1 & $\geq$1 \\
\bottomrule[1.2pt]
\end{tabular}
\begin{tablenotes}
    \scriptsize
    \item [1]User engagement indicators are weighted by Eq.~(2); at least one indicator must meet the threshold.
    \item [2]Functional capability indicators use original values; all indicators in the corresponding category must meet their thresholds.
\end{tablenotes}
\end{threeparttable}
}
\end{table}

The thresholds are determined through multiple rounds of empirical validation and data analysis. In particular, user engagement indicators are closely aligned with the specific characteristics of the LLM app store, such as platform popularity and user activity levels. Therefore, when adapting this framework to other platforms, we recommend calibrating these thresholds based on the actual data distribution of the target environment to ensure optimal evaluation performance.
Once the thresholds are established, we apply the time-weighted evaluation algorithm described in \autoref{subsubsec:time} to compute weighted user engagement scores for each LLM app. Because user engagement indicators such as views and interactions are directly influenced by operating history, \tool{} uses the time-weighted method defined in Eq.~(2)  to reduce the disadvantage faced by newly launched apps. Functional capability indicators, by contrast, are not affected by operating history and are therefore evaluated using their original values. During screening, an app is retained if at least one weighted user engagement indicator meets its threshold, whereas all functional capability indicators in the corresponding category must satisfy their thresholds.

At this stage, we intentionally adopt relatively inclusive thresholds to efficiently filter out apps with clear functional shortcomings, while minimizing the risk of excluding promising candidates. This strategy helps maintain a sufficiently large and high-quality candidate pool for subsequent, more detailed dynamic evaluations.

\subsection{Dynamic Scenario-Adaptive Evaluation}
\label{subsec:dynamic}

Dynamic scenario-adaptive evaluation is the core of \tool{}, which addresses the limitations of traditional fixed metrics by dynamically constructing evaluation criteria for a variety of application scenarios. Based on the labels generated for each LLM app by an LLM-based annotator (\autoref{subsec:tag}), the system systematically designs relevant evaluation metrics and scoring criteria. It then generates corresponding evaluation tasks, ensuring that the quantitative evaluation of LLM apps accurately reflects their real-world functions and usage contexts.

\subsubsection{Evaluation Metric and Task Generation}
\label{subsubsec:generation}
\begin{figure}[htbp]
    \centering
    \includegraphics[width=\linewidth]{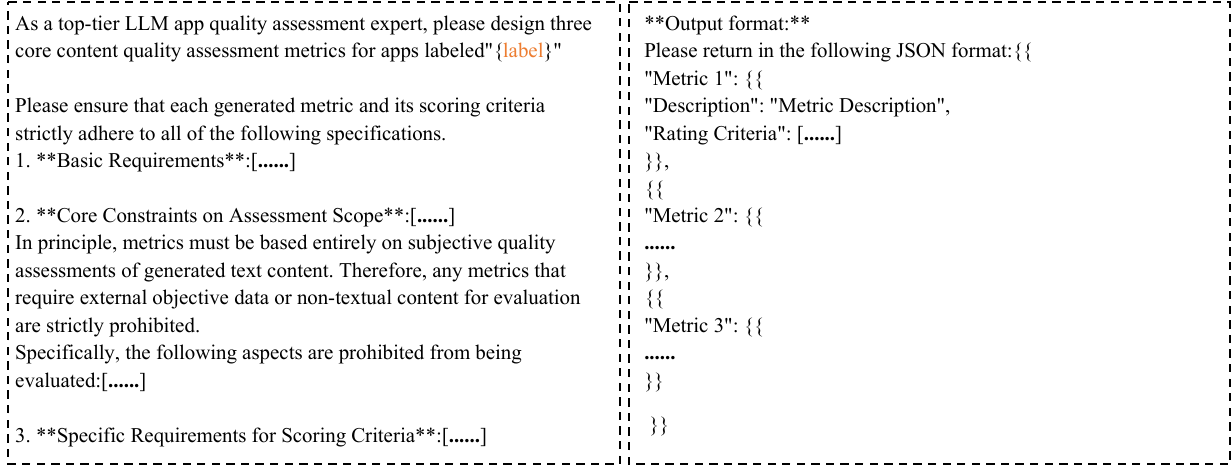}
    \caption{Prompt Template for Dynamic Evaluation Metric Generation.}
    \label{fig:prompt_2}
\end{figure}

For each LLM app, the label generated by the LLM-based annotator (\autoref{subsec:tag}) serves as the anchor for designing targeted evaluation metrics. Guided by the principles of functional uniqueness, quantifiability, and content orientation, \tool{} constructs a set of metrics that directly correspond to the app's core capabilities. The prompt template shown in Figure~\ref{fig:prompt_2} is used to instruct the LLM to generate evaluation metrics based on the given label, resulting in three core metrics for each label. Corresponding scoring criteria are also generated for each evaluation metric. By providing segmented descriptions for each score level, these metrics are translated into clearly structured and quantifiable scoring guidelines.

\begin{figure}[htbp]
    \centering
    \includegraphics[width=\linewidth]{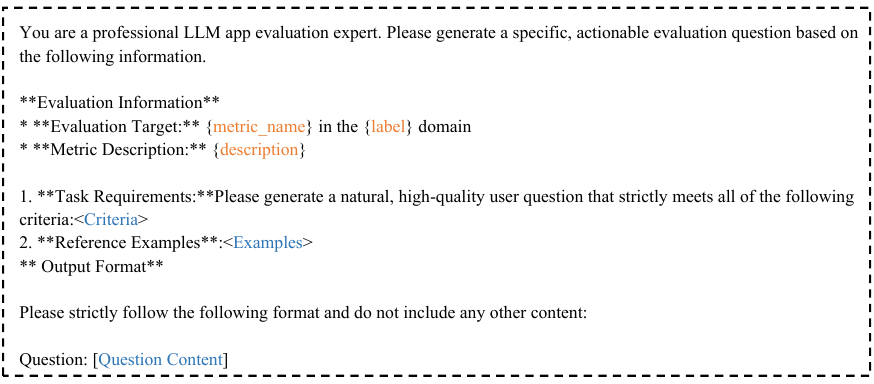}
    \caption{Prompt Template for Dynamic Evaluation Tasks Generation.}
    \label{fig:prompt_3}
\end{figure}

Based on these metrics, \tool{} dynamically generates evaluation tasks tailored to each LLM app. Leveraging the core label and corresponding metrics, \tool{} simulates real user interaction to create tasks that reflect authentic user needs and diverse usage contexts. \tool{} can produce detailed and targeted evaluation tasks, with multiple rounds of quality verification to ensure both accuracy and reliability.
As illustrated in Figure~\ref{fig:prompt_3}, each evaluation prompt integrates the metric name and description, explicit task requirements, instructive examples, and relevant content constraints. This ensures that every task is well-aligned with specific evaluation objectives.

To guarantee task quality, \tool{} employs a two-stage filtering process. First, syntactic integrity inspection is performed to eliminate structurally flawed tasks, such as those with incomplete sentences, incorrect punctuation, or ambiguous wording. Second, 
complexity inspection assesses whether the tasks contain sufficient information, context, and challenge to enable effective evaluation of the target metric. Tasks that are overly simplistic, vague, or lacking in necessary detail are filtered out at this stage. If a candidate task fails either check, a backoff and retry mechanism is triggered: \tool{} regenerates the task, with potential adjustments to prompts or constraints, until all quality criteria are satisfied and a high-quality evaluation task is produced.
The transformation of abstract metrics into concrete, context-specific evaluation tasks enables \tool{} to establish a standardized and adaptable framework. This design supports flexible and comprehensive evaluation across a wide range of LLM apps.

\subsubsection{Response Quality Evaluation.}

The current response quality evaluation focuses on text-centric LLM apps whose primary outputs are textual responses. Multimodal LLM apps involving image, audio, or video generation are not considered in this evaluation. Building on the evaluation metrics and evaluation tasks generated in \autoref{subsubsec:generation}, we next focus on systematically evaluating the quality of the dynamic interactions between \tool{} and LLM apps. \tool{} integrates both \textbf{content quality evaluation} and \textbf{response performance evaluation}, forming a closed feedback loop from app evaluation to actionable improvement suggestions. Such a design provides a scientific basis for the functional iteration and performance tuning of LLM apps.

\begin{figure}[htbp]
    \centering
    \includegraphics[width=\linewidth]{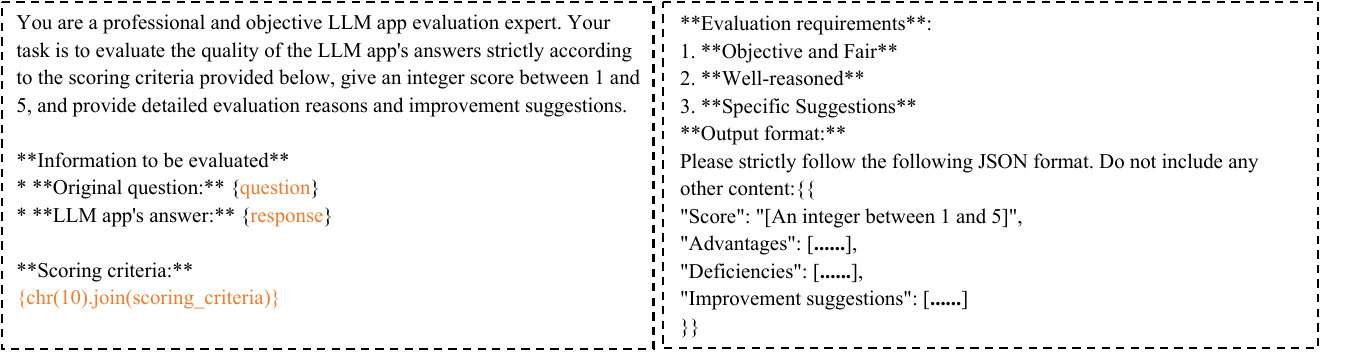}
    \caption{Prompt Template for Content Quality Evaluation.}
    \label{fig:prompt_4}
\end{figure}

\textbf{Content quality evaluation.}
This evaluation adopts an LLM-based evaluation paradigm, in which a structured prompt template guides the evaluation model to make quality judgments on LLM app responses. The prompt template, as illustrated in Figure~\ref{fig:prompt_4}, is constructed using a context embedding approach.
It explicitly incorporates three essential elements: \textit{original task}, \textit{response content}, and \textit{scoring criteria}, thus forming a logically complete evaluation context. Semantic alignment between the task and response is emphasized to ensure the evaluation model accurately understands the correspondence between user intent and generated content. Meanwhile, explicit scoring criteria transform abstract evaluation metrics into concrete judgment rules, thereby reducing subjective bias in the evaluation process.

To ensure the stability and consistency of evaluation results, model parameters are finely configured during the LLM invocation. 
Specifically, we set \texttt{temperature=0} to eliminate randomness and enhance reproducibility; \texttt{frequency\_penalty=0.5} to discourage repetitive expressions and increase informativeness; \texttt{max\_tokens=5000} to accommodate complex evaluation scenarios, and \texttt{timeout=60} seconds to prevent abnormal response delays.
Besides, five-level scale values [1, 5] are used to quantitatively rate content quality, with each level clearly defined. Alongside the score, the evaluation model outputs a structured analysis of advantages, weaknesses, and targeted improvement suggestions. The final quantified score obtained through this entire systematic evaluation process is defined as the \textit{content quality score}, denoted as $S_{CQ}$. This comprehensive framework ensures systematic and objective evaluation, providing precise guidance for app optimization.

\textbf{Response performance evaluation.}
This evaluation quantitatively analyzes the real-time interaction capability of LLM apps, with a core focus on ensuring a fluid user experience. As low latency and high throughput are critical for smooth real-time interactions, our design centers on a comprehensive performance metric: \emph{response efficiency} ($\eta$). The metric is defined as:
\begin{equation}
    \eta = \frac{\text{Tokens}}{\text{R\_time}}
\end{equation}
where \texttt{R\_time} denotes the total elapsed time (in seconds) from the user's query submission to the receipt of the last token in the LLM-generated response, and \texttt{Tokens} refers to the total number of tokens in the response.
By collecting real-time performance data during LLM app operation, response efficiency provides a direct measure of the system’s speed and capability in processing and generating outputs. To translate this continuous metric into intuitive and comparable scores, we adopt a discretization strategy based on predefined efficiency thresholds. 

Specifically, a lookup table, as shown in Table~\ref{tab:response_efficiency_scoring}, is designed to map each measured response efficiency value to a categorical score within [1, 5]. The final quantified score obtained through this method is defined as the \textit{response performance score},
denoted as $S_{RP}$. This module offers a quantifiable basis for performance diagnosis, resource allocation optimization, and service quality assurance, substantially enhancing the practical and engineering value of the overall evaluation framework.

\begin{table}[h!]
  \centering
  \caption{Response Efficiency Scoring.}
  \label{tab:response_efficiency_scoring}
  \resizebox{1\linewidth}{!}{%
    \begin{tabular}{lccccc}
      \toprule[1.2pt]
      \textbf{Metric Dimension (Quarterly)} & \textbf{Excellent (5 pts)} & \textbf{Good (4 pts)} & \textbf{Average (3 pts)} & \textbf{Poor (2 pts)} & \textbf{Very Poor (1 pt)} \\
      \midrule[1.2pt]
      \textbf{Response Efficiency(tokens/sec)} & $\ge$25 & $\ge$20 & $\ge$15 & $\ge$10 & $<$10 \\
      \bottomrule[1.2pt]
    \end{tabular}%
  }
\end{table}

\subsubsection{Weighted Composite Scoring}

After obtaining the independent \textit{content quality score} ($S_{CQ}$) and \textit{response performance score} ($S_{RP}$), a composite scoring mechanism is applied to ensure that the evaluation results comprehensively and objectively reflect the overall quality of LLM apps. The final composite score is calculated as follows:
\begin{equation}
\text{Score} = \alpha \times S_{CQ} + \lambda \times S_{RP}
\end{equation}
where the weighting coefficients satisfy $\alpha + \lambda = 1$ and $\alpha >  \lambda$. Based on empirical analysis, $\alpha$ is set to $0.8$ and $\lambda$ is set to $0.2$. 
The reason is, although response performance is important for user experience, its marginal utility is generally lower than that of content quality. By prioritizing content quality in the weighting, the evaluation system encourages the optimization of substantive output while still considering improvements in response efficiency.
The composite evaluation score serves not only as a quantitative indicator of overall application performance, but also as an important basis for iterative optimization. This comprehensive evaluation enables quick identification of both the overall performance level and the specific strengths or weaknesses of an LLM app, providing robust decision support for the continuous improvement and targeted tuning of LLM apps.

\section{Evaluation}
\label{sec:evaluation}

This section systematically examines the effectiveness of the \tool{} framework in real-world LLM app scenarios. We focus on evaluation accuracy, recommendation rationality, and user experience. 
Our evaluation is guided by the following research questions (RQs): 

\noindent\hangindent=2.5em\hangafter=1\textbf{RQ1: Can \tool{} provide accurate and scenario-adaptive quality evaluation for various types of LLM apps?}
This question aims to examine whether the automated evaluation results from \tool{} are consistent with user judgments and dynamically adapt to different LLM app scenarios.

\noindent\hangindent=2.5em\hangafter=1\textbf{RQ2: How does \tool{} compare to recommendation mechanisms in mainstream LLM app stores in terms of recommendation quality and rationality?}
Here, we investigate whether \tool{} can offer a more comprehensive and reasonable recommendation mechanism that helps users identify truly high-quality LLM apps, surpassing existing LLM app store recommendation mechanisms.

\noindent\hangindent=2.5em\hangafter=1\textbf{RQ3: What is the impact of \tool{} on user decision support, cognitive effort, and trust in real-world LLM app discovery tasks?}
We focus on determining if the adoption of \tool{} leads to concrete improvements in user decision-making, reduces mental workload, and enhances trust during the LLM app selection process.

\subsection{RQ1: Accuracy and Adaptivity}

\subsubsection{Experimental Setup}

All experiments were conducted on the AppBuilder~\cite{baiduapp}, a mainstream LLM app store, which provides a diverse set of LLM apps and a rich array of functional configurations. Two distinct scenarios were selected as experimental subjects to ensure coverage of both \textit{professional} Q\&A and \textit{tool-type} LLM app categories: (1) \textit{legal consulting} and (2) \textit{travel planning}. 
For each scenario, a single scenario-specific keyword was used to retrieve candidate LLM apps from the platform, reflecting the typical user search and recommendation logic in LLM app stores. Specifically, the keyword ``lawyer'' was used for the \textit{legal consulting} scenario, which yielded 18 candidate apps, while ``smart travel planning assistant'' was used for the \textit{travel planning} scenario, yielding 32 candidate apps. These numbers represent the original set of apps to be evaluated for each scenario.

Subsequently, each set of candidate apps was independently evaluated by \tool{}. For all LLM-based evaluation modules in \tool{}, we used ``Qwen-32B''.
It is worth noting that the underlying LLM in \tool{} is modular and can be replaced with other LLMs as needed. In parallel, a panel of 28 human evaluators, including several domain experts, assessed the same LLM apps using the same metrics. All evaluators were experienced LLM app users aged 18–50, with several possessing relevant domain expertise.

\subsubsection{Automated Evaluation by \tool{}} 
The automated evaluation process in \tool{} comprises three main stages: (1) LLM app labeling and classification, (2) static indicator evaluation, and (3) dynamic scenario-adaptive evaluation.

\textbf{LLM app labeling and classification.}
Before formal evaluation, \tool{} employs an LLM-based annotator to automatically generate concise, scenario-specific labels for each app based on its name and description. These labels distill the core functionality and typical usage scenario of the app, facilitating accurate downstream classification.
Among the 18 LLM apps retrieved using the keyword ``lawyer'' in the legal consulting scenario, the annotator consistently generated the label ``law consulting analysis'' for these apps. According to the functional taxonomy, this label corresponds to the ``Law'' tag, which is categorized under the \textit{professional} Q\&A category. Similarly, for the 32 LLM apps retrieved with the keyword ``smart travel planning assistant'' in the travel planning scenario, the annotator generated the label ``travel itinerary planning''. This label maps to the ``Planning'' tag, which falls under the \textit{tool-type} category in the taxonomy.
Each app is thus mapped to a fine-grained hierarchical taxonomy, including category, subcategory, and tag. This process ensures that subsequent evaluation criteria and thresholds are precisely aligned with the app's intended usage context, thereby enabling scenario-adaptive evaluation.

\textbf{Static indicator evaluation.}
Static evaluation aims to efficiently reduce the candidate pool size and improve baseline app quality before dynamic evaluation.
For the legal consulting scenario, the initial set of 18 apps was filtered using a combination of user engagement metrics (views, interactions, favorites, copies) and functional capability metrics (plugins, knowledge bases, bulit-in models), together with category-specific thresholds and time-decay weighting. As a result, \tool{} selected 6 high-potential candidates. Similarly, in the travel planning scenario, the candidate pool was reduced from 32 to 6 apps.
Table~\ref{tab:evaluation_rq1_static} summarizes the original static indicator values for all shortlisted apps. During screening, user engagement indicators were evaluated using weighted values, whereas functional capability indicators were compared using their original values. All retained apps satisfied the screening requirements, ensuring a suitable candidate pool for subsequent quality evaluation.

\begin{table}[htbp]
\centering
{
\small
\caption{Static Indicator Values for Shortlisted LLM Apps.}
\label{tab:evaluation_rq1_static}
}
\resizebox{\linewidth}{!}{
\begin{threeparttable}
\begin{tabular}{llcccccc|cccccc}
\toprule[1.2pt]
\textbf{Category} & \textbf{Metric} & \textbf{La1}\tnote{1} & \textbf{La2} & \textbf{La3} & \textbf{La4} & \textbf{La5} & \textbf{La6} & \textbf{Ta1}\tnote{2} & \textbf{Ta2} & \textbf{Ta3} & \textbf{Ta4} & \textbf{Ta5} & \textbf{Ta6} \\
\midrule[1.2pt]
\multirow{4}{*}{User Engagement}
    & Views      & 154 & 26  & 19  & 28   & 271  & 70   & 20  & 103 & 17  & 31  & 12  & 14 \\
    & Interactions       & 425 & 157 & 78  & 234  & 15,000& 68   & 132 & 59  & 117 & 345 & 169 & 36 \\
    & Favorites  & 1   & 0   & 0   & 0    & 4    & 0    & 2   & 0   & 0   & 0   & 0   & 0 \\
    & Copies     & 0   & 1   & 0   & 0    & 0    & 0    & 0   & 1   & 0   & 1   & 1   & 0 \\
\midrule
\multirow{3}{*}{Functional Capability}
    & Plugins & 3 & 0 & 4 & 2 & 0 & 6 & 5 & 3 & 4 & 6 & 4 & 4 \\
    & Knowledge Bases    &15 & 1 & 1 & 1 & 2 & 1 & 2 & 1 & 1 & 1 & 1 & 1 \\
    & Built-in Models    & 2 & 2 & 2 & 2 & 2 & 2 & 2 & 2 & 2 & 2 & 2 & 2 \\
\bottomrule[1.2pt]
\end{tabular}
\begin{tablenotes}
    \footnotesize
    \item [1] LaX denotes apps in the \textit{legal consulting} scenario (La = Legal consulting app).
    \item [2] TaX denotes apps in the \textit{travel planning} scenario (Ta = Travel planning app).
\end{tablenotes}
\end{threeparttable}
}
\end{table}

\textbf{Dynamic scenario-adaptive evaluation.}
Following static screening, the dynamic evaluation module of \tool{} automatically adapts its metrics to the application's scenario, leveraging the previously generated labels. For every shortlisted app, the framework dynamically constructs three core evaluation metrics tailored to the scenario, and for each metric, it defines a five-level quantitative scoring metric, values set as  [1, 5], with detailed criteria for each level. Representative scenario-specific evaluation tasks are designed to ensure the assessment’s authenticity and relevance.
For the legal consulting and travel planning scenario, \tool{} establishes three core evaluation metrics for each context, as presented in Table~\ref{tab:dynamic_metrics}. The scoring metrics ensure that different levels of app performance are clearly and fairly distinguished.

\begin{table}[htbp]
    \centering
    \caption{Scenario-Adaptive Dynamic Evaluation Metrics and Tasks in \tool{}.}
    \label{tab:dynamic_metrics}
    \resizebox{1\linewidth}{!}{
    \begin{tabular}{cl>{\raggedright\arraybackslash}p{5.8cm}>{\raggedright\arraybackslash}p{6.2cm}c}
        \toprule[1.2pt]
        \textbf{Scenario} & \textbf{Metric} & \textbf{Definition} & \textbf{Task Example} & \textbf{Score} \\
        \midrule[1.2pt]
        \multirow{11}{*}{\textbf{\shortstack{Legal\\Consulting}}}
            & \multirow{3}{*}{Legal Citation Accuracy}
                & Assesses whether legal terms cited in the answer are accurate and relevant to the question.
                & ``If an employee is dismissed without notice, which labor law articles apply?'' & \multirow{3}{*}{1--5} \\ \cmidrule{2-5}
   
            & \multirow{4}{*}{Terminology Clarity}
                & Evaluates whether legal terminology is used appropriately and clearly explained.
                & ``Explain the meaning of `force majeure' in a contract dispute to a layperson.'' & \multirow{3}{*}{1--5} \\ \cmidrule{2-5}

            & \multirow{4}{*}{Answer Completeness \& Logic}
                & Judges if the answer addresses key legal points, is logically organized, and has a well-founded conclusion.
                & ``I'm a tenant and I need to break my lease early. What are the legal steps I need to follow? What risks should I be aware of?'' & \multirow{3}{*}{1--5} \\ 

        \midrule
        \multirow{11}{*}{\textbf{\shortstack{Travel\\Planning}}}
            & \multirow{4}{*}{Personalization Match}
                & Measures whether the plan meets all personalized requirements (e.g., budget, interests, time constraints).
                & ``Plan a 3-day trip to Chengdu for a family with kids, with a budget under 3000 RMB.'' & \multirow{3}{*}{1--5} \\ \cmidrule{2-5}
 
            & \multirow{4}{*}{Itinerary Logic \& Flow}
                & Assesses the reasonableness and flow of the itinerary, including transportation and activity planning.
                & ``Design a 5-day Japan itinerary covering Tokyo, Kyoto, and Osaka with minimal backtracking.'' & \multirow{3}{*}{1--5} \\ \cmidrule{2-5}
       
            & \multirow{3}{*}{Content Innovation}
                & Evaluates whether recommendations include unique and insightful local attractions or activities.
                & ``Recommend lesser-known attractions or local experiences in Florence for art lovers.'' & \multirow{3}{*}{1--5} \\ 

        \bottomrule[1.2pt]
    \end{tabular}
    }
\end{table}

LLM apps demonstrated varying performance across core metrics. In the legal consulting scenario, most apps received scores of 3–4, indicating generally good performance, while a few had lower scores on certain metrics, such as legal citation accuracy. In the travel planning scenario, top apps consistently scored higher, while others showed weaknesses in itinerary logic and personalization. These results highlight the framework’s ability to provide fine-grained, scenario-relevant evaluation and to identify both strengths and weaknesses of each app.

\subsubsection{Human Evaluation and Consistency Analysis}
To benchmark the accuracy and adaptivity of \tool{}, the six shortlisted apps from each scenario were independently evaluated by the panel of 28 human raters, including domain experts, using the same dynamically generated metrics and a 5-point scale. We then compared the scores from \tool{} with the average human ratings.
\begin{figure}[htbp]
    \centering
    \begin{subfigure}{0.49\linewidth} 
        \includegraphics[width=\linewidth]{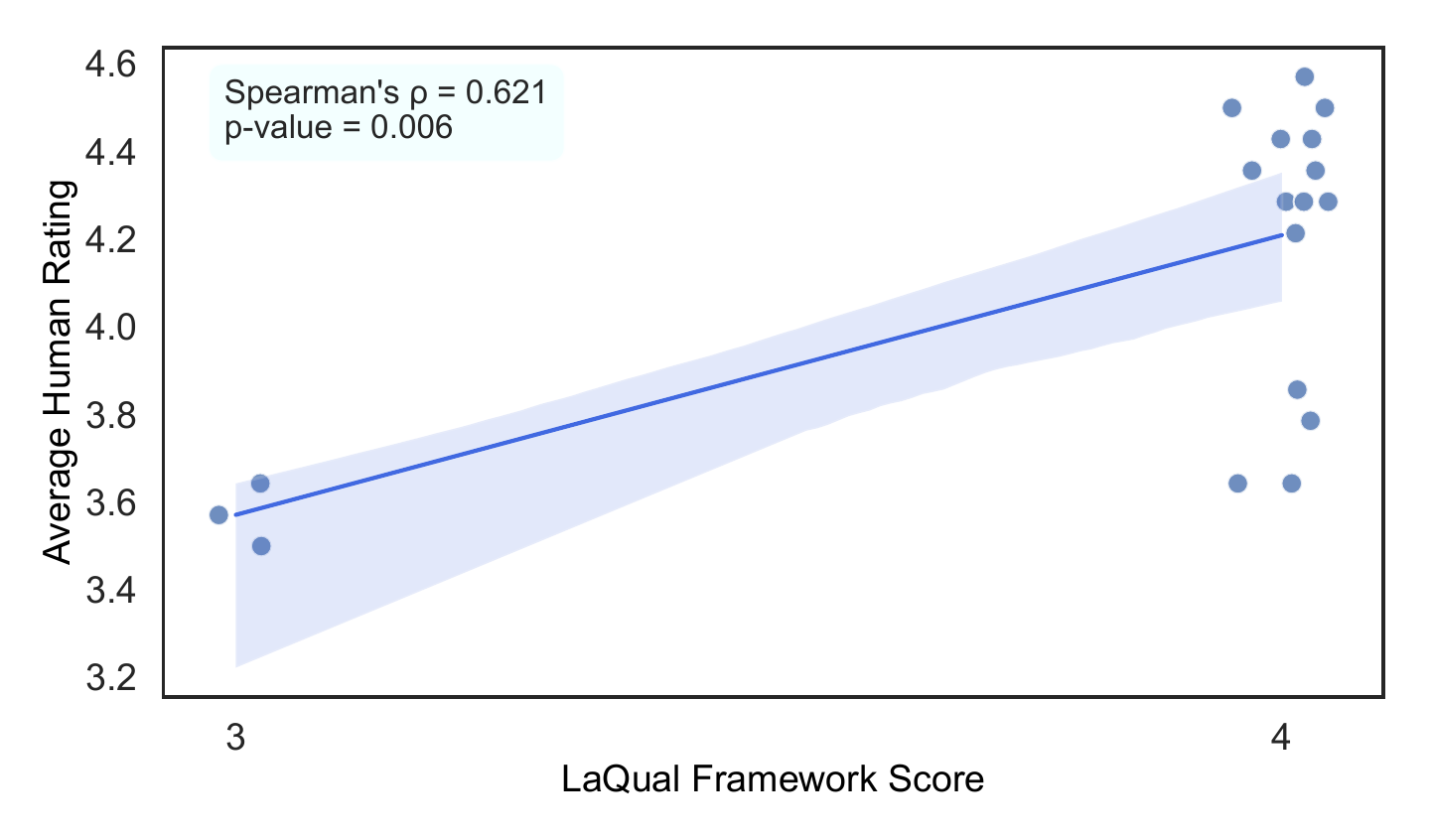}
        \caption{\textit{Legal Consulting} Scenario.}
        \label{fig:law}
    \end{subfigure}
    \hfill
    \begin{subfigure}{0.49\linewidth}
        \includegraphics[width=\linewidth]{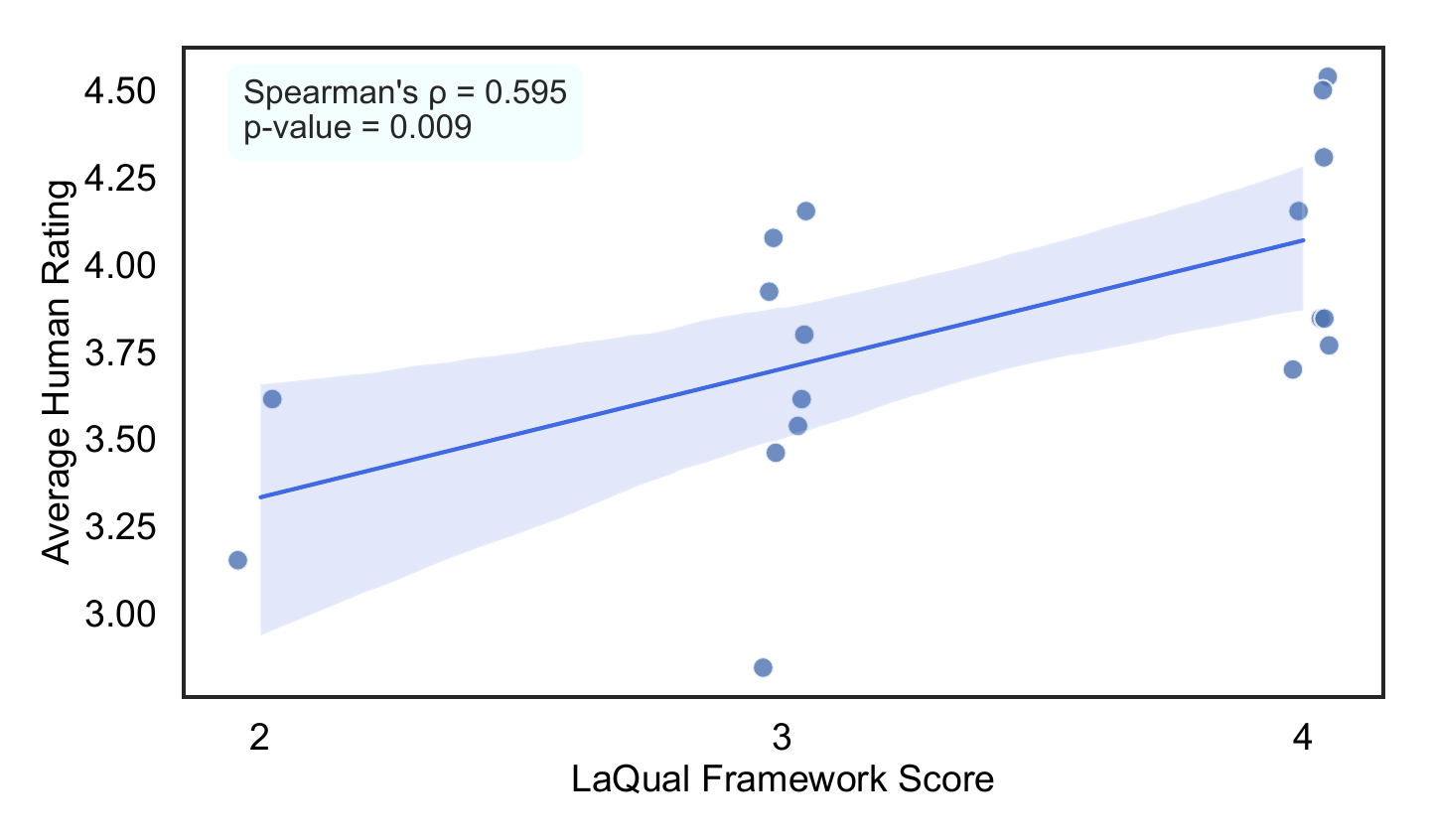}
        \caption{\textit{Travel Planning} Scenario.}
        \label{fig:travel}
    \end{subfigure}
    \caption{Correlation Analysis between \tool{} and Human Evaluation.}
\end{figure}

For the legal consulting scenario, Spearman's rank correlation analysis was conducted between the human average scores and \tool{}'s content scores across all 18 evaluation items (six apps, three metrics each). As shown in Figure~\ref{fig:law}, the correlation coefficient ($\rho$)~\cite{schober2018correlation} was \textbf{0.621} with a p-value of \textbf{0.006}, indicating a moderately strong and statistically significant positive correlation. This demonstrates that \tool{}'s automated evaluation is highly consistent with human expert judgment.
Regarding response performance, it's noted that no direct correlation with human ratings was assessed due to differences in measurement: human scores reflect subjective fluency perceptions, while \tool{} computes objective response efficiency. Nevertheless, the response performance score ($S_{RP}$) remains a critical component of the final composite evaluation, ensuring that both content quality and interaction fluency are reflected in the overall assessment.

Similarly, for the travel planning scenario, the consistency between \tool{} and human raters was evaluated using the same method. As shown in Figure~\ref{fig:travel}, Spearman's correlation coefficient was \textbf{0.595} with a p-value of \textbf{0.009}, again indicating a moderately strong and statistically significant positive correlation. This further validates \tool{}'s adaptability and accuracy across different application domains.

\begin{shaded}
\noindent \textbf{Answer to RQ1:}  
The results show that \tool{} achieves high consistency with human judgments across diverse LLM app scenarios. In the legal consulting case, the Spearman correlation coefficient between \tool{}’s automated scores and user ratings reached 0.621 (p = 0.006); in the travel planning case, the coefficient was 0.595 (p = 0.009). This indicates that \tool{} not only provides accurate evaluations, but also dynamically adapts its assessment criteria to fit different domains.
\end{shaded}

\subsection{RQ2: Recommendation Effectiveness}

RQ2 conducts a further analysis based on the experimental results of RQ1 to examine whether \tool{} can provide more effective and rational recommendations than mainstream LLM app store recommendation mechanisms.

\textbf{Manual verification procedure and consensus.} To better understand the exclusion patterns produced by the static screening stage, we conducted a qualitative verification of the 38 excluded apps based on consensus among the volunteers (12 in the legal scenario and 26 in the travel planning scenario). This verification was carried out by three volunteers with relevant experience in software quality research. Specifically, the verification was conducted in two phases: (1) \textit{Independent Examination}, in which each volunteer independently examined the apps' metadata, platform indicators, and functional configurations; and (2) \textit{Consensus Discussion}, in which the team collectively discussed the results to reach agreement on the major exclusion patterns. The verification suggests that the excluded apps consistently exhibited the following recurring characteristics:

\begin{itemize}[noitemsep, leftmargin=*]
\item \textbf{Lack of market vitality:} zombie apps with very low or declining user activity, or early-stage apps that had not yet gained meaningful user recognition.
\item \textbf{Lack of domain support:} apps that relied mainly on foundation model knowledge and lacked integrated knowledge bases or other support relevant to the scenario.
\item \textbf{Functional deficiencies:} especially in the travel planning scenario, apps that lacked essential real-time tool integrations (e.g., map navigation and weather plugins).
\item \textbf{Superficial design:} apps whose functions or interaction design showed limited differentiation and failed to address the core needs of the target scenario.
\end{itemize}

\textbf{Comparison with the default app store ranking.} We further compared \tool{}'s selected apps with the store's default \textit{top-6 most popular} results. We observed a partial mismatch between the store’s popularity based ranking and \tool{}’s screening results: \tool{} retained only 3 of the store's Top-6 apps in the legal scenario and 2 of the store's Top-6 apps in the travel planning scenario.
Among the highly ranked platform apps that were still filtered out, we observed recurring limitations in support relevant to the scenario. In particular, a common pattern was the lack of plugins or knowledge base support needed for the scenario. In some cases, these apps also exhibited imbalanced engagement signals, such as single-digit page views paired with hundreds or even thousands of interactions. By contrast, several apps retained by \tool{} but not ranked in the platform’s default Top-6 showed more complete functional configurations than some of the highly ranked apps that were filtered out. These results suggest that \tool{} does not simply reproduce popularity-based ranking, but helps identify apps that are more consistent with scenario-specific requirements.

\begin{shaded}
\noindent \textbf{Answer to RQ2:}  
\tool{} can provide more effective and rational recommendations than mainstream LLM app store recommendation mechanisms. For example, in the legal consulting scenario, \tool{} reduced the candidate pool from 18 to 6 apps (a 66.7\% reduction), and in the travel planning scenario, from 32 to 6 apps (an 81.25\% reduction), by systematically filtering out apps that did not satisfy the user engagement or functional requirements of the framework. Manual verification further suggests that the excluded apps typically lacked sufficient market vitality, domain support, or essential functional capabilities, indicating that \tool{} captures recommendation cues beyond popularity alone and helps identify apps that are better aligned with the requirements of the scenario.
\end{shaded}

\subsection{RQ3: User Impact}

This RQ examines the practical value of the framework in real-world LLM app discovery scenarios. Specifically, it investigates whether the dynamic evaluation mechanism provided by \tool{} can substantially improve decision support, cognitive efficiency, and trust, compared to a baseline system. The overarching goal is to assess the framework’s effectiveness as a decision support tool for LLM app selection.

\subsubsection{Research Hypotheses}

Based on the core challenges users face in the LLM app store, such as information overload and difficulty discerning quality, we propose the following hypotheses:
\begin{itemize}[itemsep=2pt, topsep=3pt, left=2pt]
\item \textbf{RH1: Decision support advantage.} One system may better help users identify apps that match their core needs, leading to more confident final choices.
\item \textbf{RH2: Cognitive efficiency advantage.} One system may reduce users’ mental effort during app comparison and selection, thereby improving comparison efficiency and decision confidence.
\item \textbf{RH3: Trust and transparency advantage.} One system may be perceived as more transparent, credible, and helpful for supporting app discovery decisions.
\end{itemize}

\subsubsection{Task Design}

To empirically validate these hypotheses, we conducted a controlled user study with a within-subjects design. Twelve participants were recruited to use both the \tool{} framework and a baseline system (i.e., AppBuilder) for app discovery and selection under the same application scenario and decision goal. The study focused on two application scenarios related to legal consulting and travel planning. To mitigate order effects, participants were randomly assigned to two groups in a counterbalanced sequence:

\begin{description}
\item[Group A (n=6):] Used the baseline system first, then \tool{}.
\item[Group B (n=6):] Used \tool{} first, then the baseline system.
\end{description}

For each scenario-based app discovery task, participants entered scenario-relevant keywords in both conditions. Examples included ``lawyer'' and ``civil law'' for the legal scenario, and ``travel itinerary planning'' and ``travel guide'' for the travel scenario. Participants were allowed to choose their own keywords within each predefined scenario, as users may have different needs even within the same broad scenario. In the baseline condition, participants entered the keyword into the live AppBuilder platform, browsed the apps returned by the default search and ranking mechanism, and identified suitable apps. The displayed information included app names, descriptions, rankings, and engagement information provided by the platform. In the \tool{} condition, participants entered the same keyword into the \tool{} console, reviewed the apps recommended by \tool{}, and made their selections based on the generated evaluation report. For each recommended app, \tool{} provided a direct trial URL, the overall evaluation score, scenario-specific metric scores, and the corresponding advantages, deficiencies, and improvement suggestions for each evaluation metric.

\begin{figure}[htbp]
    \centering
    \includegraphics[width=1\linewidth]{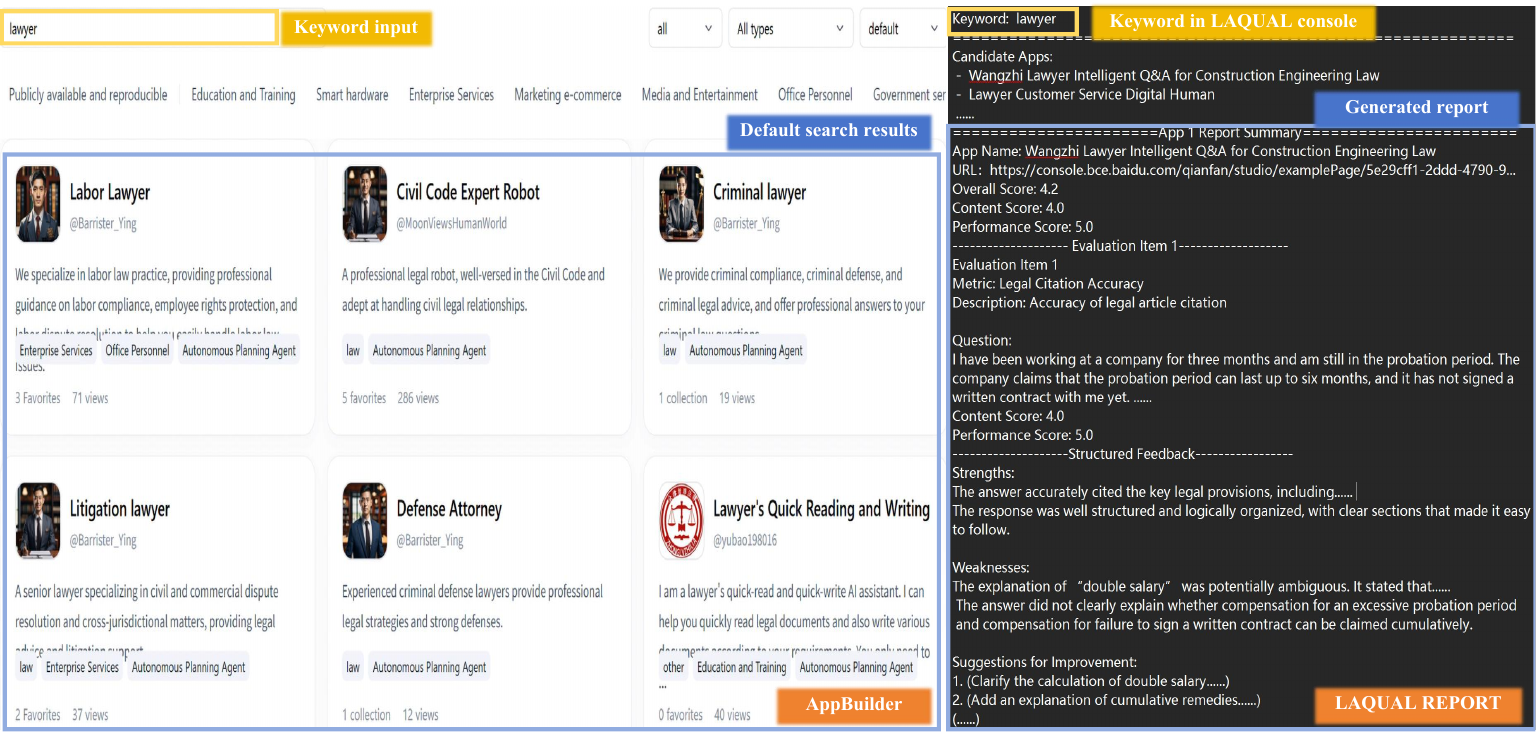}
    \caption{Representative excerpts from the baseline AppBuilder results and the recommendation report generated by \tool{}.}
    \label{fig:user_study_interfaces}
\end{figure}

Although the exact keywords varied across participants, each participant used the same keyword in both conditions for the same task. This within-subject design allowed the two systems to be compared under the same user need for each participant. The study was conducted through direct interaction with the live platform and \tool{}. Figure~\ref{fig:user_study_interfaces} shows representative interface excerpts from the two conditions: the baseline search-result interface and the recommendation report generated by \tool{} after the same keyword was entered.

After completing all tasks, participants filled out a post-task questionnaire based on a seven-point Likert scale (1 = Strongly Disagree, 7 = Strongly Agree). The questionnaire was designed around the three research hypotheses and covered six evaluation dimensions, as detailed in Table~\ref{tab:user_study_metrics}.

\begin{table}[htbp]
    \centering
    \caption{Quantitative Evaluation Dimensions and Metrics for the User Study.}
    \label{tab:user_study_metrics}
    \begin{threeparttable}
    \resizebox{\linewidth}{!}{
    \begin{tabular}{l c l p{7.8cm}}
        \toprule[1.2pt]
        \textbf{Dimension} & \textbf{Weight} & \textbf{Metric} & \textbf{Description} \\
        \midrule[1.2pt]
        \multirow{6}{*}{\textbf{Decision Support Benefits}}
            & \multirow{3}{*}{15\%} & \multirow{3}{*}{Decision Confidence} & The information provided by the system\tnote{1} helped me make a final choice with confidence. \\ \cmidrule{2-4}
            & \multirow{3}{*}{15\%} & \multirow{3}{*}{Comparison Efficiency} & The system helped me compare different LLM apps more efficiently, reducing my cognitive and decision-making load. \\
        \midrule
        \multirow{5}{*}{\textbf{Evaluation Quality \& Credibility}}
            & \multirow{3}{*}{15\% }& \multirow{3}{*}{Value of Explanatory Information} & The information and evidence provided by the system were substantively helpful for my decision-making. \\ \cmidrule{2-4}
            & \multirow{2}{*}{15\%} & \multirow{2}{*}{Result Credibility} & I found the results provided by the system objective and reliable. \\
        \midrule
        \multirow{3}{*}{\textbf{Depth of Demand Matching}}
            & \multirow{3}{*}{20\%} & \multirow{3}{*}{Deep Demand Fulfillment} & The system provided results that addressed both my immediate needs and relevant considerations. \\
        \midrule
        \multirow{2}{*}{\textbf{Exploration \& Discovery Value}}
            & \multirow{2}{*}{20\%} & \multirow{2}{*}{Discovery of ``Surprise''} & The system helped me identify additional app options worth considering. \\
        \bottomrule[1.2pt]
    \end{tabular}
    }
    \begin{tablenotes}
    \scriptsize
    \item [1] ``The system'' refers to either \tool{} or Baidu Qianfan.
    \end{tablenotes}
    \end{threeparttable}
\end{table}

\subsubsection{Results} 
We analyzed the questionnaire data using a paired-samples t-test, as each participant experienced both systems. The results, summarized in Table~\ref{tab:t_test_results}, indicate that \tool{} performed better than the baseline system across all reported evaluation dimensions.

\begin{figure}[htbp]
    \centering
    \includegraphics[width=0.75\linewidth]{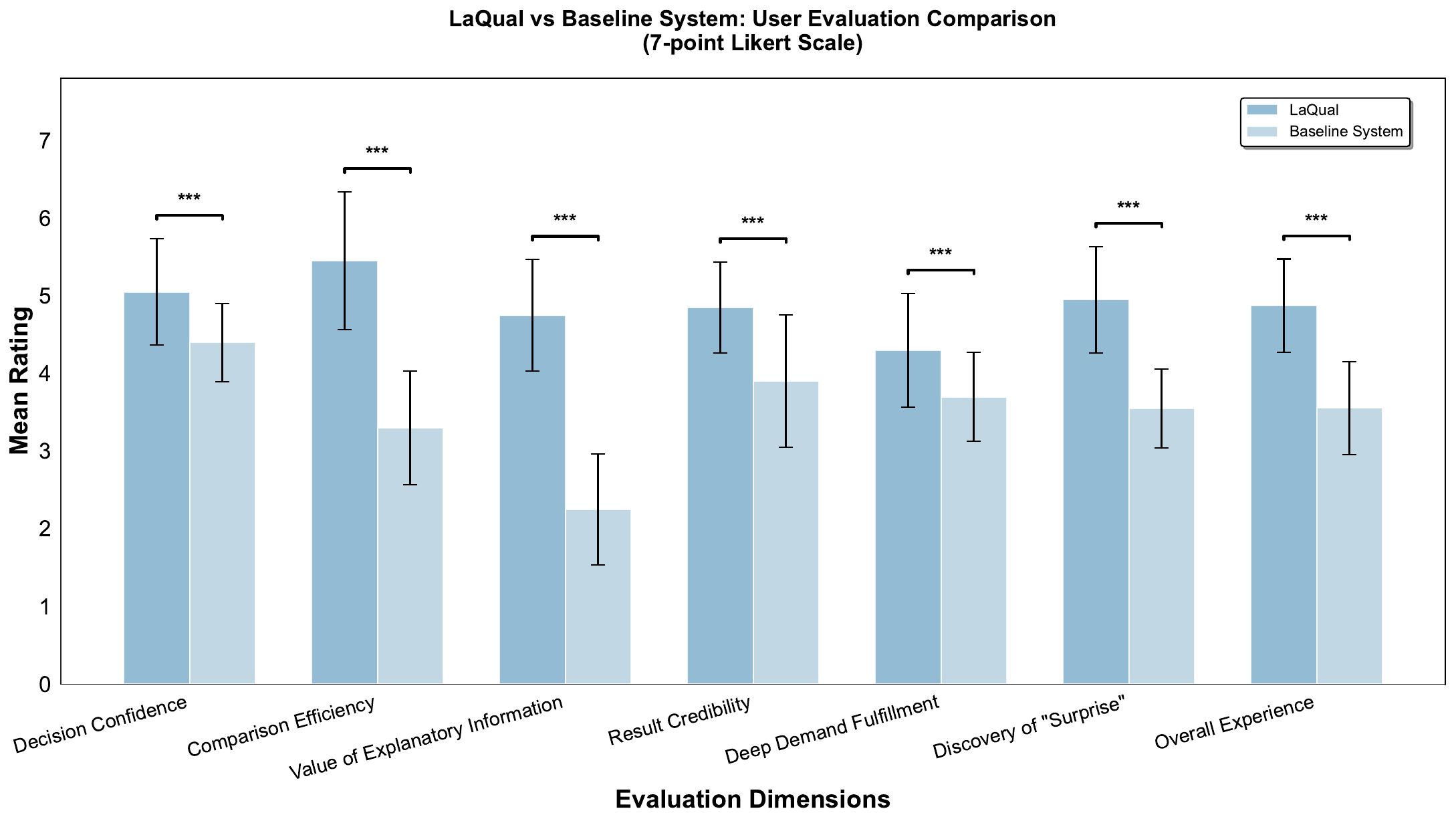}
    \caption{Mean Score Comparison of \tool{} and the Baseline System Across All User Evaluation Dimensions.}
    \label{fig: Mean Score Comparison}
\end{figure}

As visualized in  
Figure~\ref{fig: Mean Score Comparison}, \tool{} achieved notably higher mean scores (all statistically significant at the p < 0.001 level, as indicated by *** in the figure). The advantage was particularly pronounced in the dimensions of \textit{comparison efficiency} (M=5.45 vs. M=3.30) and \textit{Value of Explanatory Information} (M=4.75 vs. M=2.25). These results suggest that the information provided by \tool{} helps users make app comparisons more efficiently and confidently.
 
Furthermore, \tool{} also showed higher scores in \textit{decision confidence}, \textit{result credibility}, and the \textit{discovery of ``surprise''}, further indicating its positive effects on user confidence and trust.

\begin{table}[htbp]
    \centering
    \caption{Paired-Samples T-Test Results Comparing \tool{} and the Baseline System.}
    \label{tab:t_test_results}
    \resizebox{1\linewidth}{!}{
    \begin{tabular}{llcccc}
        \toprule[1.2pt]
        \textbf{Evaluation Category} & \textbf{Sub-dimension} & \textbf{\tool{} Mean} & \textbf{Baseline Mean} & \textbf{\tool{} SD} & \textbf{Baseline SD} \\
        \midrule[1.2pt]
        \multirow{2}{*}{Decision Support Benefits}
            & Decision Confidence (Dim. 1) & 5.050 & 4.400 & 0.686 & 0.503 \\
            & Comparison Efficiency (Dim. 2) & 5.450 & 3.300 & 0.887 & 0.733 \\
        \midrule
        \multirow{2}{*}{Evaluation Quality \& Credibility}
            & Value of Explanatory Information (Dim. 3) & 4.750 & 2.250 & 0.716 & 0.716 \\
            & Result Credibility (Dim. 4) & 4.850 & 3.900 & 0.587 & 0.852 \\
        \midrule
        Depth of Demand Matching
            & Deep Demand Fulfillment (Dim. 5) & 4.300 & 3.700 & 0.733 & 0.571 \\
        \midrule
        Exploration \& Discovery Value
            & Discovery of ``Surprise'' (Dim. 6) & 4.950 & 3.550 & 0.686 & 0.510 \\
        \midrule
        \textbf{Weighted Total Score}
            & \textbf{Overall Experience Score} & \textbf{4.873} & \textbf{3.555} & - & - \\
        \bottomrule[1.2pt]
    \end{tabular}
    }
\end{table}

\begin{shaded}
\noindent \textbf{Answer to RQ3:}  
User studies indicate that \tool{} improves users' decision support, reduces cognitive effort, and enhances trust. Quantitative results show that \tool{} achieved higher mean scores than the baseline system across all reported evaluation dimensions, with particularly strong advantages in comparison efficiency (M=5.45 vs. M=3.30) and value of explanatory information (M=4.75 vs. M=2.25). These findings suggest that \tool{} contributes to more confident, efficient, and trustworthy LLM app selection experiences.
\end{shaded}

\section{Implications}
\label{sec:implication}
Our evaluation framework provides actionable benefits and concrete guidance for all key stakeholders in the LLM app ecosystem, supporting the establishment of a more transparent, high-quality, and trustworthy environment.

\underline{\textbf{LLM app store managers.}}
For store managers, our findings demonstrate that relying solely on static metrics such as downloads or user ratings can easily result in ranking manipulation and poor user experience. By integrating \tool{} into routine platform operations, managers can implement automated quality evaluations at both the onboarding and periodic review stages. For example, only apps that meet minimum quality thresholds in scenario-adaptive evaluation can be listed, and regular re-evaluations can help identify apps with declining performance for timely intervention or removal. Additionally, publishing composite and detailed evaluation scores alongside traditional popularity indicators allows users to make more informed decisions, while also encouraging developers to focus on genuine quality improvements.

\underline{\textbf{LLM app developers.}}
For developers, \tool{} offers a set of actionable, scenario-specific quality metrics that can be integrated directly into development and release workflows. Developers can use these evaluations during prototyping to quickly identify weaknesses in content, functionality, or user experience, and address them before public release. Incorporating \tool{} into continuous integration pipelines enables automatic quality checks with each update, reducing the risk of quality regressions. Furthermore, detailed evaluation reports can help developers prioritize improvements in areas that matter most to users, such as response accuracy or feature completeness. By transparently sharing evaluation results, developers can also strengthen user trust and better position high-quality LLM apps in a competitive market.

\underline{\textbf{LLM app users.}}
End users benefit from our framework through greater transparency and confidence in the LLM apps they select. With access to multi-dimensional evaluation results, users can choose apps that best match their individual needs. For instance, prioritizing response accuracy in legal apps or customization in travel apps. The availability of sub-scores and scenario-based explanations also makes it easier for users to understand the strengths and weaknesses of each app, reducing decision fatigue and the risk of disappointment. As a result, users are empowered to make more informed and satisfying choices, and can provide valuable feedback to further refine quality standards.

\underline{\textbf{Researchers.}}  
For the research community, this research contributes two core concepts to the field of LLM app quality evaluation. The first is the adaptability of the evaluation model. We revealed that the inherent judgment logic of the \tool{} evaluation framework varies depending on the attributes of the metric. Although all evaluations ultimately output a unified five-point scale, the framework's internal logic for deriving this score shifts based on the inherent differences in the app scenario. When evaluating metrics with clear objective criteria or knowledge boundaries, such as ``accuracy of legal references'' the criteria are highly quasi-binary, making qualitative grading easier. However, when evaluating metrics that essentially measure ``completeness'' or ``match'' such as ``personalized match of travel plans'' the focus is more on the completeness of the multi-functional package or personalized satisfaction, resulting in a wider distribution of final scores. We believe that it is precisely through this flexibility in its inherent judgment logic that \tool{} achieves accurate evaluation across scenarios.

Secondly, we reveal that tool{} acts not just as an evaluator, but also as a \textbf{stabilizer} for the evaluation ecosystem. It should be considered a stabilizer within the evaluation ecosystem. A recurring phenomenon in our empirical research is that the ratings from human evaluation panels often exhibit substantial dispersion, reflecting significant differences between evaluators due to subjective experience and preferences. In contrast, \tool{} provides an objective and stable rating for app evaluations, acting as a stabilizer.

\section{Threats to Validity}
\label{sec:limitation}
Despite the comprehensive and systematic design of the \tool{} framework, several threats to validity remain. We recognize these limitations and have taken targeted measures to mitigate their impact where possible. This section discusses threats from the perspectives of content, internal, and external validity.

\underline{\textbf{Content validity.}}
In the static indicator evaluation, content validity may be affected by the subjectivity of threshold setting. Since complete platform data is unavailable, the admission threshold in this framework is determined by practical experience, and its optimality requires further validation. Additionally, fixed thresholds may inadvertently exclude apps targeting niche scenarios. To minimize these threats, we adopt a low-threshold filtering strategy to first eliminate unqualified applications, reducing the risk of filtering out potential high-quality apps. We also set differentiated functional indicator requirements for different app categories. While this approach is preliminary, it considers the characteristics of various apps and maintains filtering efficiency, enabling the framework to achieve scenario-based and accurate evaluation while retaining overall generality.

\underline{\textbf{Internal validity.}}
The internal validity of our evaluation may be affected by inherent bias within LLMs. Since the evaluation process is powered by a LLM, it is susceptible to political, cultural, or social biases present in its training data or architecture~\cite{cao2023assessing, bai2024measuring}. These biases can influence scoring results, particularly in subjective or open-ended scenarios, and may compromise fairness and objectivity. To mitigate this risk, several strategies are employed. First, we use carefully designed prompt instructions to constrain LLM behavior. Second, we adopt standardized and scenario-adaptive scoring metrics to minimize ambiguity. Third, we set deterministic decoding parameters, such as temperature at zero, to enhance the consistency of results. Despite these measures, it is not possible to fully eliminate model-internal bias, and interpretation of evaluation outcomes should take this limitation into account.

\underline{\textbf{External validity.}}

Several factors may affect the generalizability of our findings. The reported evaluation covered two scenarios: 18 LLM apps for legal consulting and 32 LLM apps for travel planning. These scenarios include both professional Q\&A and tool-type apps, and the results should be interpreted as evidence from a focused empirical study. The evaluation was conducted on Baidu AppBuilder. Because LLM app stores may differ in ranking mechanisms, metadata structures, and user engagement indicators, some components of \tool{}, such as static thresholds, may need to be calibrated before the framework is used on other platforms. In addition, the evaluation used concise search terms to reflect common app discovery behavior. Future work should consider more diverse and complex query settings to further test the robustness of the framework. The experiments were conducted in the Chinese platform environment of Baidu AppBuilder, using Chinese prompts and app content during the actual evaluation. For readability and consistency, the examples in this paper are presented in English. The framework itself is not tied to a specific language; when \tool{} is applied to platforms in other linguistic environments, the prompts, taxonomy terms, and scenario descriptions can be adapted to the target language.

In addition, the current \tool{} is limited to evaluating text-based outputs of LLM apps. As the LLM app ecosystem is evolving toward multi-modal applications, including those involving image generation and audio-visual content, our evaluation methods may not be directly applicable to these new forms. To partially mitigate these limitations, we designed our evaluation tasks and metrics to be scenario-adaptive and structurally extensible, providing a foundation for future adaptation to multi-modal or hybrid applications. However, the current conclusions should be interpreted within the scope of text-centric LLM app scenarios, and extending the framework to multi-modal evaluation remains an important direction for future research.

\section{Related Work}
\label{sec:related}

\subsection{Research on LLM Apps}

The explosive growth of LLM apps has given rise to a novel software ecosystem centered on LLM app stores. Although academic research on this ecosystem is still in its infancy, significant strides have already been made in several key areas. Current studies primarily concentrate on three dimensions: ecosystem analysis, security vulnerabilities, and data privacy. 

At the ecosystem level, pioneering efforts have begun to map the landscape of this rapidly developing field. Zhao et al.~\cite{zhao2025llm} laid the groundwork by proposing the first comprehensive research roadmap for the LLM app ecosystem. Foundational datasets and initial exploratory analyses were contributed by Hou et al.~\cite{Hou2024GPTZoo}, who constructed the large-scale GPTZoo dataset, and by Yan et al.~\cite{Yan2024Exploring}, who investigated patterns in LLM app distribution.
In terms of security, Hou et al.~\cite{Hou2025On} were among the first to expose the security threats facing LLM app stores, systematically identifying several attack vectors such as app hijacking and prompt injection. Building on this, Xie et al.~\cite{xie2025llm} conducted an in-depth analysis of app squatting and cloning, highlighting the severe risks these imitation attacks pose to the overall trust and stability of the ecosystem.
Regarding data privacy, researchers have revealed that the flexibility inherent in app configuration can also introduce significant risks of data leakage. Li et al.~\cite{yan2025understanding} demonstrated that private application knowledge files are susceptible to theft via prompt engineering, while Wu et al.~\cite{wu2024an} uncovered the widespread issue of excessive user data collection throughout the ecosystem.
While existing research has made notable advances in understanding the ecosystem’s structure as well as its security and privacy risks, most efforts remain focused on external threats and vulnerabilities. Little attention has been paid to systematically evaluating the intrinsic quality of content generated by LLM apps. This represents an important research gap. The framework proposed in this study, \tool{}, seeks to fill this gap by enabling systematic and automated evaluation of LLM app content quality.

\subsection{Domain-Specific LLM Evaluation}

A growing body of research has focused on evaluating LLMs within specialized domains, resulting in a variety of tailored frameworks and benchmarks that address the unique requirements and challenges of different professional fields.
In the healthcare domain, recent work has addressed both the opportunities and challenges of applying LLMs to medical tasks. Krolik et al.~\cite{krolik2024towards} explored the use of LLMs for automated medical Q\&A, while Nazi et al.~\cite{nazi2024large} argued that evaluation should consider not only generative ability but also practical applicability and safety in real-world scenarios. Building on this, Ei et al.~\cite{wei2024evaluation} introduced a dedicated evaluation framework for medical queries, finding that ChatGPT achieves high accuracy in this context.
In finance, Guo et al.~\cite{guo2023chatgpt} proposed FinLMEval to systematically evaluate LLMs on a range of financial tasks. Xie et al.~\cite{xie2024finben} further developed FinBen, the first comprehensive benchmark for financial LLM evaluation, enabling more detailed assessment of model performance in complex financial domains.
In education, Xiao et al.~\cite{xiao2023evaluating} built a system for generating reading comprehension exercises using LLMs, validating its effectiveness through both automatic and manual evaluation. Kavadella et al.~\cite{kavadella2024evaluation} examined ChatGPT's role in undergraduate dental education, showing improved student performance with model assistance.
The legal field has also seen robust evaluation efforts. Wang et al.~\cite{wang2024legal} compared LLM-generated legal content to reference materials and expert judgments. Li et al.~\cite{li2024legalagentbench} presented LegalAgentBench, which uses metrics such as success rate, progress rate, and BERT-Score for hierarchical legal tasks. Haitao Li et al.~\cite{li2024lexeval} introduced LexEval for systematic evaluation in the Chinese legal context, while Savelka et al.~\cite{savelka2023unlocking} showed that GPT-3.5 achieves high accuracy in zero-shot legal text annotation.

These domain-specific evaluation frameworks underscore the substantial promise of LLMs in professional and expert-driven contexts. However, their scope is typically confined to individual scenarios, lacking the scalability and adaptability required for broader application landscapes. This limitation highlights a central motivation for our work: to move beyond fixed-scenario evaluations and develop a universal, automated framework capable of dynamically generating scenario-aware evaluation criteria for LLM apps.

\subsection{Metrics for LLM evaluation}
\label{secTraditional evaluation indicators for LLM evaluation}
Mainstream evaluation of LLM content quality largely relies on traditional natural language processing metrics. Surface-level semantic matching is typically assessed using BLEU~\cite{papineni2002bleu} and ROUGE~\cite{lin2004rouge}, which measure text similarity based on n-gram statistics such as word order and vocabulary overlap. METEOR~\cite{banerjee2005meteor} improves upon these by enhancing the ability to capture semantic equivalence between texts. More recently, metrics like BERTScore~\cite{zhang2019bertscore} and BLEURT~\cite{sellam2020bleurt} have advanced the field by leveraging pre-trained language models to assess deep semantic similarity, representing a shift from pure surface matching to semantic-level evaluation.
However, these methods still exhibit notable limitations in the context of LLM evaluation. Specifically, they often lack domain adaptation and deep semantic understanding, making it difficult to accurately assess outputs that involve creative expression or specialized domain knowledge. As a result, there can be significant discrepancies between automated evaluation scores and the actual semantic quality perceived by users in real-world LLM applications.

\subsection{Evaluation Based on Human Preferences}
While traditional algorithm-based evaluation methods offer efficiency, they often struggle to capture the nuanced preferences and subjective judgments of human users. Belz and Reiter~\cite{dong2022survey} provided a comprehensive review of human evaluation techniques for natural language generation, underscoring the indispensable value of human judgment in assessing generated text quality. To better align models with user expectations, Ouyang et al.~\cite{ouyang2022training} introduced reinforcement learning from human feedback (RLHF), where models are trained using human preference rankings for different outputs. Similarly, Chatbot Arena~\cite{chiang2024chatbot} employs crowdsourcing to evaluate LLMs based on large-scale human preferences, while Shankar et al.~\cite{shankar2024validates} proposed EvalGen, a hybrid framework aiming to better align evaluators with human expectations. In the context of code generation, Yang et al.~\cite{yang2024evaluating} highlighted the critical importance of model alignment with human preferences through systematic evaluation.
Although these approaches mark a shift from traditional semantic matching to preference-based assessment, they generally require extensive human annotation and face challenges in maintaining annotation consistency. Consequently, their scalability and practicality for automated, large-scale evaluation in LLM app ecosystems remain limited.

\section{Conclusion} 
\label{sec:conclusion}
In this work, we present \tool{}, a unified and automated framework for large-scale LLM app quality evaluation. Our approach systematically integrates scenario-aware hierarchical classification, multi-dimensional static indicators filtering, and LLM-driven dynamic evaluation to provide fine-grained and context-adaptive assessment across diverse LLM app scenarios. Extensive experiments on real-world app store data demonstrate that \tool{} achieves strong alignment with expert judgments, substantially reduces the candidate app pool, and significantly improves user decision quality, efficiency, and trust compared to existing recommendation mechanisms. Furthermore, our framework is highly extensible and can be adapted to new evaluation scenarios as the LLM app ecosystem evolves.
Moving forward, we plan to further enhance the interpretability of automated evaluation results and explore integration with real-time user feedback, aiming to further enhance the robustness and trustworthiness of LLM app discovery experiences.

\section*{CRediT authorship contribution statement}
\textbf{Yan Wang:} Writing - original draft, Writing - review \& editing, Software, Investigation, Formal analysis, Methodology, Visualization. 

\textbf{Xinyi Hou:} Writing - review \& editing, Methodology, Validation, Investigation, Visualization. 

\textbf{Junjun Si:} Writing - review \& editing, Supervision, Methodology, Investigation.

\textbf{Yanjie Zhao:} Writing - review \& editing, Methodology, Validation. 

\textbf{Weiguo Lin:} Writing - review \& editing, Methodology, Supervision. 

\textbf{Haoyu Wang:}  Writing - review \& editing, Supervision, Resources, Methodology. 

\section*{Declaration of generative AI and AI-assisted technologies in the manuscript preparation process}

During the preparation of this work, the authors used ChatGPT in order to improve the English language clarity and grammar of the manuscript. After using this tool, the authors reviewed and edited the content as needed and take full responsibility for the content of the published article.

\section*{Data Availability}

The implementation details and evaluation results for \tool{} are publicly available at \url{https://github.com/wangyanainn-ainn/LaQual}.

\section*{Acknowledgements}

The authors would like to thank our colleagues and peers at our respective institutions for their constant support and insightful discussions. We also sincerely appreciate the volunteers and domain experts who participated in the user studies and consistency assessments. This work was sponsored by the State Key Laboratory of Multimedia Information Processing Open Fund and the Fundamental Research Funds for the Central Universities under Grant CUC25XT02.

\bibliographystyle{elsarticle-num} 
\bibliography{refs_revised}

\end{document}